# STATE EQUATION FOR DENSE GASES AND LIQUIDS FROM A SELF-CONSISTENT-FIELD APPROACH

**D. Giusti, V. Molinari, D. Mostacci[*]**

*Laboratorio di Ingegneria Nucleare di Montecuccolino*

*Via dei Colli 16, I-40136 Bologna (ITALY)*

*Alma Mater Studiorum - Università di Bologna (Bologna, Italy)*

**ABSTRACT**

The departure from ideal gas behavior is described by several known equations of state (EoS), developed from a combination of theoretical considerations and experimental correlations. In this work a different approach is proposed, in which from a pairwise potential of the Lennard-Jones type, interaction between molecules is accounted for by means of a self-consistent force field. An EoS is thus derived and compared to the Virial and to the van der Waals EoS.

## I. INTRODUCTION

When a gas is not too dense, experimental p-V isotherms may be fitted over a very wide range of temperature T and pressure p, with a power series relation

$$\frac{pV}{RT} = 1 + \frac{B(T)}{V} + \frac{C(T)}{V^2} + \frac{D(T)}{V^3} + \cdots \qquad (1)$$

---

[*] Corresponding author - fax: +39-051-6441747; tel.: +39-051-6441711; e-mail: *domiziano.mostacci@unibo.it*

which is referred to as the *virial equation of state* [1,2,3]. The temperature dependent coefficients $B(T), C(T), D(T), \ldots$ are called the second, third, fourth, ... virial coefficients. These coefficients express the deviations from ideal behavior when collisions involving two, three, four, ... molecules become important in the gas.

Thus, when the density is low, deviations from ideality are adequately described by the second term, the second virial coefficient, whereas, as densities grow higher more and more terms are needed. This can be seen also from another point of view: if $N$ is the number of molecules and if only two molecules are close to each other and they are far from the remaining $N-2$ molecules, and this is the case for all pairs in the assembly, the second virial coefficient alone is adequate to represent the deviations from ideality. If molecules are close to each other in groups of three, the third virial coefficient comes into play to account for the three-body interactions, and so forth. At very high density, as is the case of liquids, the terms in the expansion reach much higher orders; in fact, many molecules become members of very large clusters. In other words, more and more virial coefficients in the power series in Eq. (1) must be considered and this procedure becomes extremely awkward. Consequently, when the density grows large the virial expansion becomes of little or no use.

Aim of this work is to obtain an equation of state for a gas of very high density, or for a liquid, i.e., situations where every molecule is in the force field of many others, so that multiple encounters are the normal situation: molecules of the system are "clustered" together in large numbers.

This result will be obtained by means of the self-consistent field as defined in the Boltzmann-Vlasov equation [4,5]. This self-consistent field becomes important to describe the dynamic behaviour of a system where every molecule interacts simultaneously with a large number of other molecules, and correlation between pairs of molecules can be in turn disregarded since its effect becomes negligible.

From this point of view, the interactions are treated in a way reminiscent of a common practice of plasma physics, where a specific parameter - the "plasma parameter" - establishes when there are enough particles in the so-called "Debye sphere" that the self-consistent field becomes a valid description since simultaneous multiple collisions prevail [4,5].

In Section 2 the self-consistent field will be derived for the one-dimensional case assuming a Lennard-Jones-type, pair-wise potential model, modified to account also for the finite size of the molecules by means of a distance of closest approach.

In Section 3 an "equivalent pressure" is defined and then EoS for liquids and dense gases is obtained.

Finally, Section 4 compares the EoS to classical results and discusses the EoS with reference to the liquid-vapor transitions.

**II. SELF-CONSISTENT FIELD**

In the Vlasov equation [4,5]

$$\frac{\partial f}{\partial t} + \mathbf{v} \cdot \frac{\partial f}{\partial \mathbf{r}} + \frac{\mathbf{F}+\mathbf{F'}}{m} \cdot \frac{\partial f}{\partial \mathbf{v}} = 0 \qquad (2)$$

the effect of molecule interaction is accounted through a self-consistent field $\mathbf{F}'$, to be calculated as

$$\mathbf{F'}(\mathbf{r}) = \int_V n(\mathbf{r}_2)\mathbf{F}_{1,2}d\mathbf{r}_2 \qquad (3)$$

where $\mathbf{F}_{1,2}$ is the force that a molecule located at position $\mathbf{r}_2$ exerts on the molecule in $\mathbf{r}$ and $n(\mathbf{r})$ is the local number density at $\mathbf{r}$. It is worth recalling that the assumption that the total potential energy of the system may be expressed as the sum of the potentials between all pairs of molecules is basic to the derivation of eq. (3).

The detailed form of the interaction function can be investigated only through quantum mechanics and much work has been done in this direction [1,2]. However the problem is very complex and many effects are involved; moreover the structure of the molecules is often not very well known. Therefore the existing results contain significant approximations and are applicable only to specific situations. This being the case, it becomes essential to resort to a phenomenological potential $\varphi_{1,2}$.

In this work, the following Lennard-Jones model, modified to include a distance of closest approach $\sigma$ to account for the non-vanishing dimensions of the molecules, will be used to calculate the self-consistent field. The intermolecular potential $\varphi_{1,2}(r)$ (henceforth referred to as modified Lennard-Jones model or mLJ) is presented in fig.1, and is given by

$$\varphi_{1,2}(r) = \begin{cases} \infty & \text{for } r \leq \sigma \leq r_0 \\ 4\varepsilon\left[\left(\dfrac{r_0}{r}\right)^{12} - \left(\dfrac{r_0}{r}\right)^{6}\right] & \text{for } r > \sigma \end{cases} \qquad (4)$$

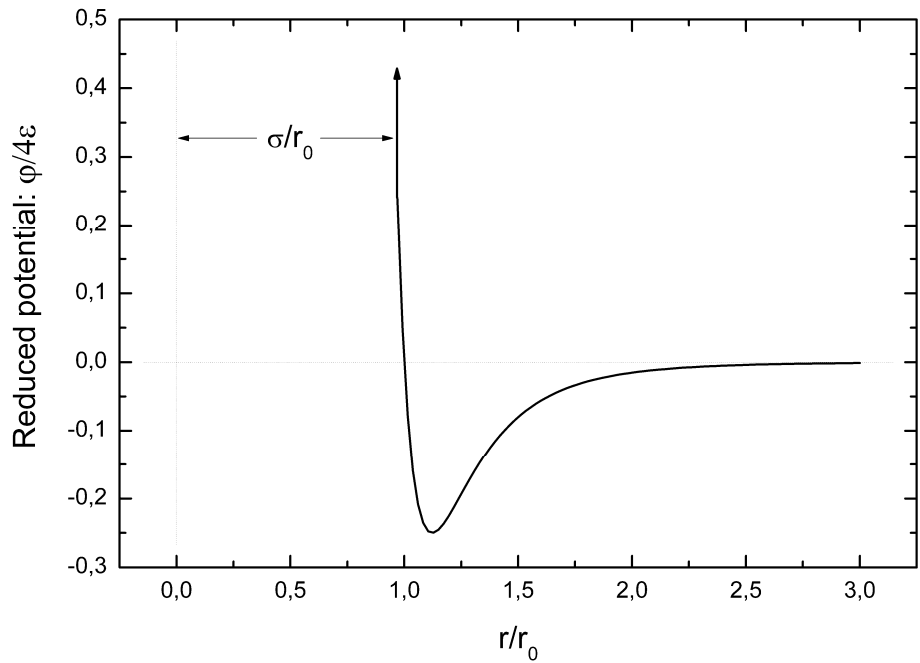

**Figure 1**: behaviour of the modified L-J potential with intermolecular distance

To calculate the self-consistent force **F'**, a molecule located at the point $(0,0,z)$ will be considered, and the force exerted on this by the whole surrounding liquid will be calculated from the mLJ potential.

To simplify the problem, a system possessing slab symmetry will be assumed, that is one in which density depends only on the z - coordinate. Consider then an elementary volume dV at a location defined by the coordinates $(r,\vartheta,\beta)$ in a spherical reference system centered in the molecule of interest and with the polar axis along the $z$ direction, see Fig. 2 [6]

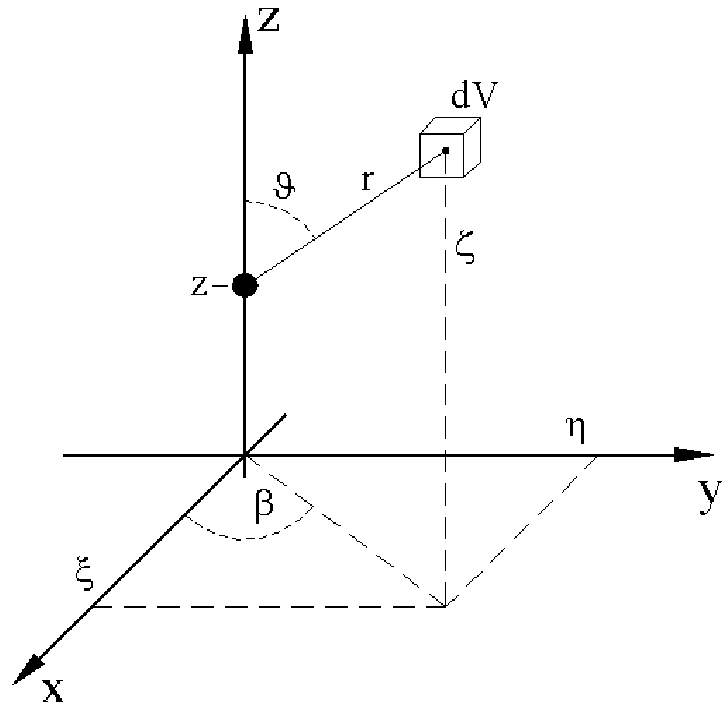

**Figure 2**: geometry of the setting

With the geometry in Fig. 2 the force acting on the molecule of interest due to a molecule in $(r,\vartheta,\beta)$ becomes

$$\mathbf{F}_{1,2} = 4\varepsilon\left[\frac{6r_0^6}{r^7} - \frac{12r_0^{12}}{r^{13}}\right]\hat{\mathbf{r}} \quad \text{for } r > \sigma \tag{5}$$

Now, calling $(\xi,\eta,\zeta)$ the cartesian coordinates of volume $dV$, the value of $r^{\alpha+1}$ with $\alpha=12$ or $\alpha=6$ can be calculated as

$$r^{\alpha+1} = \left[\xi^2 + \eta^2 + (\zeta-z)^2\right]^{\frac{\alpha+1}{2}} \tag{6}$$

Sines and cosines of the angles in Fig. 2 can be expressed in terms of the cartesian coordinates $(\xi,\eta,\zeta)$

$$\begin{cases} \sin\vartheta = \dfrac{\sqrt{\xi^2+\eta^2}}{\sqrt{\xi^2+\eta^2+(\zeta-z)^2}} & \sin\beta = \dfrac{\eta}{\sqrt{\xi^2+\eta^2}} \\ \cos\vartheta = \dfrac{\zeta-z}{\sqrt{\xi^2+\eta^2+(\zeta-z)^2}} & \cos\beta = \dfrac{\xi}{\sqrt{\xi^2+\eta^2}} \end{cases} \tag{7}$$

The Cartesian component of the force may be rewritten as

$$F_{1x} = 4\varepsilon\, n(\zeta)d\xi d\eta d\zeta \left[\dfrac{\xi r_0^\alpha \alpha}{\left[\xi^2+\eta^2+(\zeta-z)^2\right]^{\frac{\alpha+2}{2}}}\right]_{\alpha=12}^{\alpha=6} \tag{8}$$

$$F_{1y} = 4\varepsilon\, n(\zeta)d\xi d\eta d\zeta \left[\dfrac{\eta r_0^\alpha \alpha}{\left[\xi^2+\eta^2+(\zeta-z)^2\right]^{\frac{\alpha+2}{2}}}\right]_{\alpha=12}^{\alpha=6} \tag{9}$$

$$F_{1z} = 4\varepsilon\, n(\zeta)d\xi d\eta d\zeta \left[\dfrac{(\zeta-z) r_0^\alpha \alpha}{\left[\xi^2+\eta^2+(\zeta-z)^2\right]^{\frac{\alpha+2}{2}}}\right]_{\alpha=12}^{\alpha=6} \tag{10}$$

To obtain the overall force on the reference molecule, integration over the whole volume is performed. It can be seen readily that

$$\int_{-\infty}^{+\infty}\int_{-\infty}^{+\infty} F_{1x}\,d\xi\,d\eta = \int_{-\infty}^{+\infty}\int_{-\infty}^{+\infty} F_{1y}\,d\xi\,d\eta = 0 \tag{11}$$

so that there are no x and y components to the force – and this is consistent with the symmetry of the problem. As for the z component

$$\int_{-\infty}^{+\infty}\int_{-\infty}^{+\infty} F_{lz} d\xi d\eta = n(\zeta) d\zeta (\zeta - z) \left[ \int_{-\infty}^{+\infty}\int_{-\infty}^{+\infty} \frac{4\varepsilon\alpha r_0^\alpha}{\left[\xi^2 + \eta^2 + (\zeta - z)^2\right]^{\frac{\alpha+2}{2}}} d\xi d\eta \right]_{\alpha=12}^{\alpha=6} \quad (12)$$

Considering the mLJ potential (4), there is a minimum approach distance $\sigma$

$$F_L = 8\pi\varepsilon \left[ -r_0^\alpha \int_{-\infty}^{z-\sigma} \frac{n(\zeta)}{(z-\zeta)^5} d\zeta + r_0^\alpha \int_{z+\sigma}^{+\infty} \frac{n(\zeta)}{(\zeta - z)^5} d\zeta \right]_{\alpha=12}^{\alpha=6} \quad (13)$$

If the density variation is mild, $n(z)$ can be expanded in Taylor series retaining only the first few terms

$$n(\zeta) = n(z) + \frac{dn(z)}{dz}(\zeta - z) + \frac{d^2n(z)}{dz^2}\frac{(\zeta-z)^2}{2} + \frac{d^3n(z)}{dz^3}\frac{(\zeta-z)^3}{3!} + \frac{d^3n(z)}{dz^3}\frac{(\zeta-z)^4}{4!} + O\left[(\zeta-z)^5\right] \quad (14)$$

Neglecting terms of order 5 and higher, and substituting into eq. (13), after some algebra the following equation is obtained:

$$F_L(z) \cong \Lambda_1 \frac{dn(z)}{dz} + \Lambda_3 \frac{d^3n(z)}{dz^3} \quad (15)$$

where the coefficients are given by

$$\Lambda_1 = \frac{16\pi\varepsilon r_0^6}{3\sigma^3}\left[1 - \frac{1}{3}\left(\frac{r_0}{\sigma}\right)^6\right] \qquad \Lambda_3 = \frac{8\pi\varepsilon r_0^6}{3\sigma}\left[1 - \frac{1}{7}\left(\frac{r_0}{\sigma}\right)^6\right] \quad (16)$$

In what follows only the first term in (15), i.e., $\Lambda_1$, will be retained, i.e.,

$$F_L(z) \cong \frac{16\pi\varepsilon r_0^6}{3\sigma^3}\left[1 - \frac{1}{3}\left(\frac{r_0}{\sigma}\right)^6\right]\frac{dn(z)}{dz} \quad (17)$$

## III. WAVES PROPAGATION IN DENSE ASSEMBLIES

The propagation of pressure waves in liquids or dense gases is conveniently studied starting from the conservation equations derived readily from the Vlasov equation [5,7]. The continuity and momentum equations, for the present one-dimensional geometry and considering only the scalar pressure p, are written as

$$\frac{\partial n}{\partial t} + \frac{\partial}{\partial z}(nw) = 0 \tag{18}$$

$$\frac{\partial w}{\partial t} + w\frac{\partial w}{\partial z} + \frac{1}{nm}\frac{\partial p}{\partial z} - \frac{1}{m}F_L = 0 \tag{19}$$

where w is the average velocity, n the particle density and $F_L$ the self-consistent field force. A point is to be addressed: having taken the view that mutual interaction and finite size of the molecules are taken into in the mLJ potential, and hence in the self consistent field, other than for the term $F_L$ particles are to be seen as behaving like an ideal gas. Within this view, it is fully appropriate to consider for the pressure the expression relevant for ideal gases:

$$p_{id} = nK_BT \tag{20}$$

with $K_B$ the Boltzmann constant. It will be denoted $p_{id}$ in the following to keep memory of this fact.

Now, at constant temperature the pressure gradient term of the previous equation can be rewritten as

$$\frac{1}{m}\frac{dp_{id}}{dz} = \frac{1}{m}\frac{dp_{id}}{dn}\frac{dn}{dz} = a^2\frac{dn}{dz} \tag{21}$$

where

$$a^2 = \frac{1}{m}\frac{dp_{id}}{dn} = \frac{K_B T}{m} \tag{22}$$

is the speed of propagation of isothermal sound waves in ideal gases.

The force $F_L$ is that due to the self-consistent field Eq. (17). With these quantities Eq. (19) rewrites as

$$\frac{\partial w}{\partial t} + w\frac{\partial w}{\partial z} + \left\{\frac{a^2}{n} - \frac{16\pi\varepsilon}{3m}\frac{r_0^6}{\sigma^3} + \frac{16\pi\varepsilon}{9m}\frac{r_0^{12}}{\sigma^9}\right\}\frac{\partial n}{\partial z} = 0 \tag{23}$$

The term between brackets can be viewed as the derivative of a quantity, which will be called provisionally $P_{eq}$, defined by the equation

$$\frac{1}{m}\frac{dP_{eq}}{dn} = \frac{1}{m}\frac{dp_{id}}{dn} - n\frac{16\pi\varepsilon}{3m}\frac{r_0^6}{\sigma^3} + n\frac{16\pi\varepsilon}{9m}\frac{r_0^{12}}{\sigma^9} \tag{24}$$

It is proposed here to interpret the quantity $P_{eq}$ defined above as a generalized pressure term, or equivalent pressure, permitting to rewrite (19) or (23) as

$$\frac{\partial w}{\partial t} + w\frac{\partial w}{\partial z} + \frac{1}{mn}\frac{dP_{eq}}{dn}\frac{\partial n}{\partial z} = 0 \tag{25}$$

which parallels the isothermal momentum equation relevant to a gas with no interactions, a ideal gas, but with pressure replaced by the "equivalent pressure".

It will be assumed further that wave phenomena are adequately described as small perturbations from a reference configuration; that is, the local average velocity w is a small perturbation and the number density n can be written as $n = n_0 + n_1$ with $n_0$ the equilibrium value and $n_1$ a small perturbation, such that $|n_1|/n_0 << 1$ [8].

This being the case, (18) and (23) can be linearized, yielding

$$\frac{\partial n_1}{\partial t} + n_0\frac{\partial v_0}{\partial z} = 0 \tag{26}$$

$$\frac{\partial v_0}{\partial t} + \left\{ \frac{a^2}{n_0} - \frac{16\pi\varepsilon}{3m} \frac{r_0^6}{\sigma^3} + \frac{16\pi\varepsilon}{9m} \frac{r_0^{12}}{\sigma^9} \right\} \frac{\partial n_1}{\partial z} = 0 \tag{27}$$

where the mLJ parameters are evaluated at $n = n_0$. The above expressions can be combined to yield a single second order equation

$$\frac{\partial^2 n_1}{\partial t^2} - \left\{ a^2 - \frac{16\pi\varepsilon n_0}{3m} \frac{r_0^6}{\sigma^3} + \frac{16\pi\varepsilon n_0}{9m} \frac{r_0^{12}}{\sigma^9} \right\} \frac{\partial^2 n_1}{\partial z^2} = 0 \tag{28}$$

or, simply

$$\frac{\partial^2 n_1}{\partial t^2} = c_0^2 \frac{\partial^2 n_1}{\partial z^2} \tag{29}$$

representing a wave with dispersion relation $\omega = \pm c_0 k$.

It thus follows that the term in brackets, to wit

$$c_0^2 = \frac{1}{mn_0} \frac{dP_{eq}}{dn} = \left\{ a^2 - \frac{16\pi\varepsilon n_0}{3m} \frac{r_0^6}{\sigma^3} + \frac{16\pi\varepsilon n_0}{9m} \frac{r_0^{12}}{\sigma^9} \right\} \tag{30}$$

is the isothermal propagation velocity of the perturbation wave in the dense gas or liquid of unperturbed density $n_0$. The equivalent pressure in the vicinity of the reference configuration is calculated as

$$P_{eq} = p_{pg} - \left\{ n_0 \frac{16\pi\varepsilon}{3} \left[ \frac{r_0^6}{\sigma^3} \right]_{n_0} - n_0 \frac{16\pi\varepsilon}{9} \left[ \frac{r_0^{12}}{\sigma^9} \right]_{n_0} \right\} \cdot n \tag{31}$$

It is worth remarking that from measuring the sound velocity it is possible, from Eq.(30), to obtain information on the equivalent pressure.

## IV. DISCUSSION

*IV.1 Comparison with classical results*

Eq. (31) can be compared to the van der Waals equation of state

$$\left(p + \frac{a}{V^2}\right)(V - b) = RT \tag{32}$$

which may be expanded as [3,9]

$$p = nKT - am^2n^2 + bKTmn^2\chi \tag{33}$$

in the approximation of $b \ll V$. The two expressions bear a noticeable resemblance. In both Eq. (31) and Eq. (33) the first term gives the pressure of an ideal gas, the third term represents the effect arising from the finite size of the molecules and gives rise to repulsion at short distances; the second term represents the effect of the molecular cohesive forces. This last effect tends to reduce, as must be expected, the pressure of the ideal gas.

Eq.(31) can also be written

$$p = n_0(KT - Bn_0) \tag{34}$$

where B is defined as

$$B = \frac{16\pi\varepsilon}{3} \frac{r_0^6}{\sigma^3}\left[1 - \frac{1}{3}\left(\frac{r_0}{\sigma}\right)^6\right] \tag{35}$$

To compare with the classical result of eq. (1), the gas constant R can be introduced and the volume per mole $V_m$ considered, yielding

$$\frac{pV_m}{RT} = 1 - \frac{B\mathcal{N}_A}{V_m^2 KT} = 1 + \frac{C(T)}{V_m^2} \tag{36}$$

where $\mathcal{N}_A$ is the Avogadro number and $C(T) = -\frac{B\mathcal{N}_A}{KT}$.

The relevant term in the virial equation (1) represents the effect of interactions between molecules, which in the present work is accounted for by the self-consistent Vlasov field. One point may be clarified: the coefficient B depends also on the specific volume, as will be discussed in the next paragraph, so that the parallel just discussed is only formal. Keeping in mind how B originates from the self-consistent field, it contains the information of all the virial terms lumped in one single quantity.

*IV.2 The L-V transition zone*

The well known p-v (pressure-specific volume) diagram, reproduced qualitatively in Fig. 3 to better illustrate the meaning of the following nomenclature, comprises four zones: the liquid state (indicated with L in Fig. 3); the superheated vapor (marked V); the transition zone L-V, i.e., the mixture of liquid and saturated vapor in equilibrium; and the states above the critical isothermal, G for gas in Fig. 3.

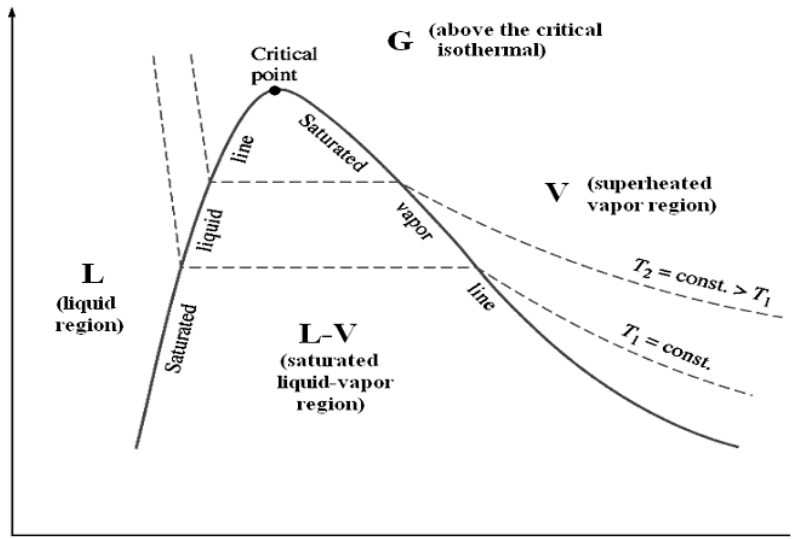

**Figure 3**: p-v diagram for a pure substance

Applying Eq.(3.16) to the zone L-V, information can be obtained on the behavior of parameter B. Let $N = N_L + N_v$ be the total number of molecules and the number of molecules in the liquid and in the saturated vapor, respectively, and let $V = V_L + V_v$ be the total volume and the volumes of the two phases. Further, $n_L = N_L/V_L$ and $n_v = N_v/V_v$ will indicate the number density of the liquid and of the vapor respectively: one must recall that these latter, in the L-V zone they are functions of temperature alone [1,10,11]. Now, from Eq.(34)

$$p = n_L(KT - B_L n_L) = n_v(KT - B_v n_v) \tag{37}$$

Then, from the value of the pressure $p(T)$ is possible to gather information about the parameter B for the two cases of the liquid and of the saturated vapor. It is to underline that, being the pressure constant in the region L-V, the parameters of the mLJ potential are different in the two phases, as it is clear from Eq.(37), since $n_L \neq n_v$ generally. Referring to the values $V_L$ and $V_G$, from Eq.(37) one gets

$$\frac{B_L}{B_G} = \frac{V_L}{V_G} \frac{pV_L/KTN - 1}{pV_G/KTN - 1} = \frac{n_v}{n_l} \frac{p/KTn_l - 1}{p/KTn_v - 1} \qquad (38)$$

Finally, B is also a function of the number density. In fact, from Eq.(37), it follows

$$B_L = \frac{1}{n_L}\left(\frac{p}{n_L} - KT\right) \; ; \; B_v = \frac{1}{n_v}\left(\frac{p}{n_v} - KT\right) \qquad (39)$$